# Exact values for the Grundy number of some graphs

Ali Mansouri[*], Mohamed salim bouhlel[**]

mehermansouri@yahoo.fr
medsalim.bouhlel@enis.rnu.tn

**Abstract:** The Grundy number of a graph G is the maximum number k of colors used to color the vertices of G such that the coloring is proper and every vertex x colored with color i, $1 \leq i \leq k$, is adjacent to (i - 1) vertices colored with each color j, $1 \leq j \leq i - 1$.
In this paper we give bounds for the Grundy number of some graphs and Cartesian products of graphs. In particular, we determine an exact value of this parameter for n-dimensional meshes and some n-dimensional toroidal meshes. Finally, we present an algorithm to generate all graphs for a given Grundy number.
**Key words:** algorithm, cartesian product of several graphs, graphs, bipartite graph, Grundy number,

## INTRODUCTION

We consider graphs without loops or multiple edges. Let G be a graph on vertices $x_0; x_1; \ldots ; X_{nG-1}$, with vertex set V (G) and edge set E (G). Let d(x) be the degree of the vertex x of G and let $\Delta$ (G) be the maximum degree of G.

The Cartesian product of two graphs G = (V1; E1) and H = (V2; E2), denoted G □H, has the vertex set $V_1 \times V_2$, and the neighborhood of each vertex $(x_1; x_2)$ is $N_{G \square H} ((x_1; x_2)) = (\{x_1\} \times N_H(x_2)) \cup (N_G(x_1) \times \{x_2\})$. Thus, in the graph G□H we find several copies of graphs G and H denoted by $G_i$ and $H_j$, where i represents the rows of G □H, with $0 \leq i \leq n_G - 1$, and j represents the columns of G □H, with $0 \leq j \leq n_G - 1$. The set $\{x_0^j; x_1^j \ldots x_{nG-1}^j\}$ denotes the vertices of the $j_h^t$ copy of G in G □H, with $0 \leq j \leq n_H - 1$.

Next, we define a k-coloring of G as a function c defined on V (G) into a set of colors

C = {1, 2,…., k} such that for each vertex $x_i$, with $0 \leq i \leq n_g - 1$, $c_{xi} \in C$. A proper k-coloring is a k-coloring verifying the condition $c_x \neq c_y$ for any pair of ad-jacent vertices x; y $\in$ V (G). A Grundy coloring is a proper k-coloring satisfying the following property P:

Every vertex v, colored with color i ($1 \leq i \leq k$), is adjacent to vertices colored by each color j such that $1 \leq j \leq i - 1$. The Grundy number $\Gamma$ (G) of a graph G is then defined as the maximum number of colors among all Grundy colorings of G. If we color only a set of vertices of a graph G, we will say that the coloring of G is partial.

This parameter was introduced by Christen and Selkow [CHS79] in 1979. They proved that determining the Grundy number is NP-complete for general graphs (also studied by McRae in 1994 [AMC94]). In [SST84], Hedetniemi et al. gave a linear algorithm for the Grundy number of a tree and established a relation between the chromatic number, the Grundy number and the achromatic number: X (G) $\leq \Gamma$ (G) $\leq \Psi$ (G), where the achromatic number $\Psi$ (G) is the maximum number of colors used for a proper coloring of G such that each pair of colors appears on at least one edge of G. In 1997, Telle and Proskurowski [GAP97] gave an algorithm for the Grundy number of partial k-trees in O (n3k2) and bounded this parameter for these graphs by the value 1 + k log2 n, where n is the graph order. In 2000, Dunbar et al. used the Grundy number to bound new parameters that they introduced in [GHK00], the chromatic and the achromatic numbers of a fall coloring. Recently, Germain and Kheddouci studied in [CGH03], [CGH06] the Grundy coloring of power graphs. They gave bounds for the Grundy number of the power graphs of a path, a cycle, a caterpillar and a complete binary tree. Such colorings are also explored for other graphs like chessboard graphs [CPR02].

The Cartesian product of graphs is widely studied in literature since it generates interesting classes of graphs like grids (obtained by the Cartesian product of paths and cycles) or hypercubes, which are used to model problems on interconnection networks or multiprocessor networks [CWE97],[SGR02], [GHW02], [FSA03]. In particular, several coloring parameters were evaluated for graphs resulting from the Cartesian product of graphs.

Cizek and Klavzar [NCK94] computed chromatic numbers of the Cartesian sum of two odd cycles. In [MK02], Kouider and Mahéo were interested in the b-chromatic number of the Cartesian product of two graphs. In [XZH92] and [XZH02], Zhu presented several bounds for respectively the star and the fractional chromatic number of graph products.

In our case, we are interested in the Grundy number of the Cartesian product of graphs. This parameter has a lot of applications in fields like scheduling or multiprocessor architectures. For instance, suppose a set of processes such that a process Pi could be computed if processes P1; P2; …; Pi-1 are already computed. Such a law on the processes can be modeled by a Grundy coloring. Thus, if we consider a given architecture G, the study of the Grundy number brings a solution for two questions.

First, how many processes can we put into this architecture? This is directly given by the Grundy number $\Gamma$ (G) of the architecture G. Second, how many times must we load processes in the architecture to compute Pn? The relation n- $\Gamma$ (G)/ $\Gamma$ (G -1 gives a solution to this second question.

Thus in this paper we will decompose our study in several parts. First in Section 2, we will present some properties of the Grundy number, by comparing it to other graph parameters. Then we will discuss in Section 3 the Grundy number of several Cartesian products of two graphs (paths, cycles, complete and bipartite graphs. . .).

In Section 4, this parameter will be studied for the Cartesian product of several graphs. Thus, we will determine an exact value of the Grundy number for n-dimensional meshes and some n-dimensional toroidal meshes. In these sections, in addition to determining exact values and bounds for the Grundy number, we will also propose constructions of these colorings. Finally in Section 5, we will present an algorithm to generate graphs for a given Grundy number

## 1. Grundy number of a graph

The First, the following obvious fact enables us to find a proper coloring for a graph G from a proper coloring of G' where G' is a subgraph of G.

**Fact 1** Let G' be an induced subgraph of G given by a set of vertices V' $\subseteq$ V. Any proper coloring of G' can be extended to a proper coloring of G.

**Proof** We extend the coloring of G' to G as follows. Let x be a vertex of G such that x $\notin$ V.' Let C be the set of colors of NG'(x). Let c be the smallest color such that c $\notin$ C. We put cx: = c and V':= V'U{x} Then, we repeat this process until V' = V.

Then, we present some results for the Grundy number of simple graphs.

**Proposition 1** Let Sn, Kn, Pn, Cn and Kn;p be respectively the stable graph, the complete graph, the path, the cycle on order n and the complete bipartite graph on n + p vertices. Let G be a non connected graph with connected components G1; G2; …; Gp. Then, we have:

**Proof.** For cases 1 to 4, the proofs are obvious.

For the complete bipartite graph, we prove the result by contradiction. Let V (Kn;p) = AUB, with A = {a1; a2; … ; an} and B = {b1; b2; … ; bp}. Let ai be a vertex colored by a color c, with $1 \leq i \leq n$ and c $\geq$ 3. There exist two vertices bj ; bk colored respectively by

| 1 .$\Gamma$ (Sn) = 1 and $\Gamma$ (Kn) = n |
|---|
| 2. $\Gamma$ (Pn) = 2 if $2 \leq n \leq 3$ <br> $\Gamma$ (Pn) = 3 if n $\geq$ 4 |
| 3. $\Gamma$ (Cn) = 2 if n = 4 <br> $\Gamma$ (Cn) =3 if n $\neq$ 4 |
| 4. $\Gamma$ (G) $\geq$ max {$\Gamma$ (Gi): $1 \leq i \leq p$} |
| 5. $\Gamma$ (Kn;p) = 2 |

colors 1 and 2, with $1 \leq j \neq k \leq p$. Observe that bk must be adjacent to color 1 on a vertex ai' , with $1 \leq i' \neq i \leq n$. So ai' and bj admit the same color 1, which is a contradiction.

Next, we show two results where the Grundy number of a graph G is bounded by other parameters of G. Firstly, we give a relation between the Grundy number and the stability number of G.

**Theorem 1** Let G be a connected graph on order n and stability number α. Then $\Gamma$ (G) $\leq$ n + 1 -α

**Proof.** Let A be an independent set in G of size α. Suppose that there is a Grundy coloring of G with k $\geq$ n - α + 2 colors. Then at least two color classes should be subsets of A. This contradicts the definition of a Grundy coloring.

Secondly, we present the inequality of "Nordhaus-Gaddum"-type for Grundy number of some graph classes.

**Proposition 2** Let G be a graph on order n and G¯ its complement. Let $\Gamma$ and $\Gamma$ ¯ be the Grundy number of G and G¯ respectively. Let x and y be two vertices of G (noted x¯ and y¯ in G¯) such that cx = $\Gamma$ and cy¯ = $\Gamma$ ¯. Then, $\Gamma$ + $\Gamma$ ¯ $\leq$ n + 1 if one of these assertions is verified:

| 1. G is a k-regular graph, with k $\geq$ 1 |
|---|
| 2. d(x) $\leq$ d(y) |
| 3. cx = $\Gamma$ and cx¯ = $\Gamma$ ¯ |

**Proof.**

**1.** The graph G is k-regular. By definition we have $\Gamma$ (G) $\leq$ ΔG) + 1 and $\Gamma$ (G¯) $\leq$ Δ (G¯) +1. Since the graph is k-regular, we deduce that Δ (G) = k and Δ (G) = n - 1 - k. Thus, $\Gamma$ (G) + $\Gamma$ (G¯) $\leq$ Δ (G) + Δ (G¯) + 2 $\leq$ n + 1.

**2.** By definition we have

d(x) $\geq$ $\Gamma$ - 1; **(1)**

And

d(y¯) $\geq \Gamma$ ¯ - 1; **(2)**

Then, inequality (2) implies

$d(y) = n - 1 - d(\overline{y}) \leq n - \overline{\Gamma}$ : (3)

As $d(x) \leq d(y)$, inequalities (1) and (3) give

$\Gamma - 1 \leq d(x) \leq d(y) \leq n - \Gamma$;

$\Gamma + \overline{\Gamma} \leq n + 1$.

**3.** From inequalities (1) and (2) we can deduce

$\Gamma + \overline{\Gamma} \leq d(x) + d(\overline{x}) + 2 = d(x) + (n - d(x) - 1) + 2 = n + 1$.

**Remark:** Observe that this inequality is not verified for any graph. Indeed, consider a graph G on order n composed by a complete graph K3 (vertices denoted x1; x2 and x3), where every node xi is the center of a star $K_{1,n_i}$, with $1 \leq i \leq 3$ and $n_i \geq 1$. By coloring G such that $c(x1) = 2$, $c(x2) = 3$, $c(x3) = 4$ and each endvertex of the stars is colored by 1 (see Figure 1.a), we find $\Gamma(G) \geq 4$. Then in the complement graph $\overline{G}$, the endvertices of the stars form a clique (denoted $K_{n1+n2+n3}$) and vertices x1, x2 and x3 are the centers of the stars respectively $K_{1;n2+n3}$, $K_{1;n1+n3}$ and $K_{1;n1+n2}$.

Therefore by coloring $c(x1) = c(x2) = c(x3) = 1$ and every vertex of the clique with a different color c, with $2 \leq c \leq n-2$ (see Figure 1.b); we have $\Gamma(\overline{G}) \geq n-2$. Thus, $\Gamma(G) + \Gamma(\overline{G}) \geq n + 2$. This remark presents a counterexample1 to the inequality of "Nordhaus-Gaddum"-type for any graph.

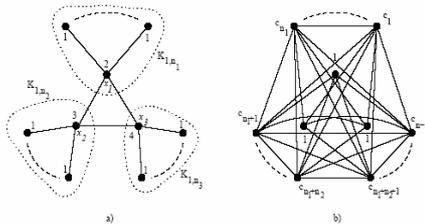

*Figure 1: Grundy colorings of a) G and b) $\overline{G}$ where $2 \leq c1 \neq c2 \neq ... \neq cn-3 \leq n-2$.*

## 2. Grundy number of the Cartesian product of two graphs

In this section, we will discuss the Grundy number of the Cartesian product of two graphs. In particular we will study the Cartesian product of two paths, two cycles, a path by a cycle, a bipartite graph by other graphs and a complete graph by any graph G.

**Proposition 3** Let G be a bipartite graph. Let Pn be a path on order $n \geq 3$ and Cm be a cycle on order $m \geq 4$. Then,

.

| $\Gamma(G \square Pn) \geq \Gamma(G) + 2$ |
|---|
| $\Gamma(G \square Cm) \geq \Gamma(G) + 2$. |

**Proof.** On the copy G1, we put the coloring of G incremented by 2. Thus, on G1 we find the colors 3 to $\Gamma(G) + 2$. Let Xi and Yi be the two independent sets of the copy Gi. We put the color 1 on every vertex of X0 and Y2, and the color 2 on every vertex of X2 and Y0. If some vertices are not colored, Fact 1 allows us to color them with a proper coloring. Thus we can deduce $\Gamma(G \square Pn) \geq \Gamma(G) + 2$ and $\Gamma(G \square Cm) \geq \Gamma(G) + 2$.

As we studied the Cartesian product of G by a path or a cycle, where G is a bipartite graph, we discuss the same products if G is not bipartite.

**Proposition 4** Let G be a non bipartite graph on order nG. Let Pn be a path on order $n \geq 4$ and Cm be a cycle on order $m \geq 4$. Then:

| $\Gamma(G \square Pn) \geq \Gamma(G) + 1$; |
|---|
| $\Gamma(G \square Cm) \geq \Gamma(G) + 1$. |

**Proof.** The proof is given by construction. We color G0; G1 and G2 with the coloring of G. Let $x_p^0$ be a vertex of G0 colored by $\Gamma(G)$. Then we put $cx_p^0 = \Gamma(G) + 1$.

$cx_p^1 = \Gamma(G)$ and $cx_p^2 = \Gamma(G) - 1$. Next we can remove the colors of vertices $x_j^1$ on G1 for which $cx_p^1 = 1$ (with $0 \leq j \neq p \leq nG - 1$) and the colors of vertices $x_i^2$ on G2 for which $cx_i^2 = \Gamma(G) - 1$ (with $0 \leq i \neq p \leq nG - 1$). Then for every remaining colored vertex on G1, we compute $cx_j^1 = cx_j^1 - 1$

Finally, by Fact 1, we color the remaining non colored vertices with a proper coloring. Thus we can deduce:

| $\Gamma(G \square Pn) \geq \Gamma(G) + 1$ and $\Gamma(G \square Cm) \geq \Gamma(G) + 1$. |
|---|

**Remark** Let Pn and Pm be the paths on order n and m respectively. And let Cn, Cm and Ck be the cycles on respectively n, m and k vertices. Then,

| **1.** $\Gamma(Pn \square Pm) = (\leq)$ 4 if n = 2 or n = m = 3; |
|---|
| 5 otherwise; |
| **2.** $\Gamma(Pn \square Ck) = $ 4 if n = 2 or n = k = 3; |
| 5 otherwise; |
| **3.** $\Gamma(Cn \square Cm) = (\leq)$ 4 if n = 3 and m = 4 or n = m = 3; |
| 5 otherwise. |

**Proof.** In each case, the maximum degree is 4. So the Grundy number is less or equal to 5. The lower bounds are obvious; they are deduced from constructions (see Figure 2).

In the following theorems we study the Grundy number for the Cartesian product of a complete graph by another graph. Firstly, we present the Cartesian product of a bipartite graph by a complete graph.

**Theorem 2** Let G be a bipartite graph on order nG. Let Kp be a complete graph on order p ≥ 3. Then,

$$\Gamma(G) + p - 1 \leq \Gamma(K_p \square G) \leq p + \Delta(G).$$

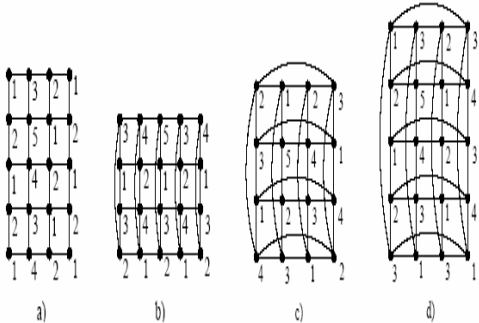

*Figure 2: Grundy colorings of a) P4 □P5, b) P5 □C4, c) C4 □C4 and d) C4 □C5.*

**Proof.** We prove by construction that $\Gamma(K_p \square G) \geq \Gamma(G) + p - 1$. Let Xi and Yi be the two independent sets of the copy Gi. We color G0 by $cx^0_i = cx_i + p - 1$ where $cx_i$ is the color of the vertex xi of G, with $0 \leq i \leq nG - 1$. Then, for each $1 \leq i \leq p - 1$, we color each vertex of Xi with the color i, and each vertex of Yi with the color i + 1 (mod (p - 1)). For example, Figure 3.b gives a Grundy coloring of K4□G. As $\Delta(K_p \square G) = \Delta(G) + p - 1$, we can deduce that $\Gamma(K_p \square G) \leq \Delta(G) + p$.

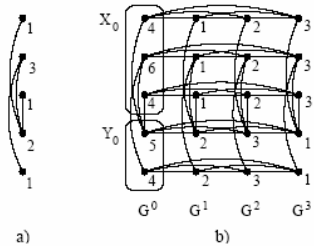

*Figure 3: Grundy colorings of a) G and b) K4 □G*

**Remark:** For all the classes of bipartite graphs such that $\Gamma(G) = \Delta(G) + 1$, the equality $\Gamma(K_p \square G) = \Gamma(G) + p - 1$ holds.

Secondly we bound the Grundy number of the Cartesian product of a complete graph by any graph G.

**Theorem 3** Let G be a graph on order nG. Let Kn be a complete graph on order n. Then,

$$\Gamma(K_n \square G) \geq \begin{cases} n + \Gamma(G) - 1 & \text{if } \Gamma(G) \leq n - 1. \\ 2n - 2 & \text{if } n \leq \Gamma(G) \leq 2n - 3. \\ \Gamma(G) & \text{if } \Gamma(G) \geq 2n - 2. \end{cases}$$

$$\Gamma(K_n \square G) \leq n + \Delta(G).$$

**Proof.** As $\Delta(K_n \square G) = \Delta(G) + n - 1$, we have $\Gamma(K_n \square G) \leq \Delta(K_n \square G) + 1 \leq n + \Delta(G)$. The proof of the lower bound is given by construction.

- $\Gamma(G) \leq n-1$. Let $k = n + \Gamma(G) - 1$. For each vertex $x^i_0$ with $0 \leq i \leq n_G - 1$, we shift the color of the vertex xi on G by (n-1)

Then, for each vertex $x^i_1$ with $0 \leq i \leq nG - 1$, we put $cx^i_1 := cx_i$. Next, for each remaining vertex, we put $cx^i_j := cx^j_{i-1} + 1 \pmod{n-1}$, with $2 \leq i \leq n-1$ and $0 \leq j \leq nG - 1$. Figure 4.b shows a Grundy coloring of K5 □G.

$n \leq \Gamma(G) \leq 2n - 3$. Let $k = 2n - 2$. Let xi, with $0 \leq i \leq nG - 1$, be a vertex of G colored by $\Gamma(G)$. As $\Gamma(G) \geq n$, then $x_i$ (resp. $x^i_0$) has at least n - 1 neighbors in G (resp. $G_0$). Let $\{v_1; v_2; ...; v_{n-1}\}$ be a set of n - 1 vertices of $N_{G_0}(x^i_0)$. To color Kn □G we put the colors (k; k - 1; .....; n) respectively on vertices $x^i_0$; $x^i_1$ ......... ; xi. Then, on each copy $K^j_n$ containing a vertex vm, with $0 \leq j \leq nG - 1$ and $1 \leq m \leq n - 1$, we put the colors (m; m + 1; .....; m + n - 2), where each color is taken modulo (n - 1), on vertices $x^j_0$; $x^j_1$ .... $x^j_{n-2}$ respectively.

- $\Gamma(G) \geq 2n - 2$. We put the coloring of G on G0 and Fact 1 completes the coloring of Kn □G to have a proper coloring.

## 3. Grundy number of the Cartesian product of several graphs

The results presented in the previous sections enable us to deduce some results for the Cartesian product of several graphs. Firstly, we determine exact values for the Grundy number of the Cartesian product of several paths and the Cartesian product of several even cycles.

**Proposition 5[BEH02]** Let Pn1; Pn2; .... ; Pnk be the paths of sizes respectively n1; n2;....; nk such that $k \geq 2$, $n_i \geq 3$ for each $1 \leq i \leq k$ and max $\{n_i : 1 \leq i \leq k\} > 3$. Let Cm1; Cm2; ...; Cmk be the cycles of sizes respectively m1; m2; .... ;mk, where $k \geq 2$, $m_i \geq 4$ and $m_i$ is even, for every $1 \leq i \leq k$. Then,

| $\Gamma(P_{n1} \square P_{n2} \square ... \square P_{nk}) = 2k + 1$ |
|---|
| $\Gamma(C_{m1} \square C_{m2} \square .... \square C_{mk}) = 2k + 1$ |

**Remark:** The graph Pn1 □Pn2 □…□Pnk is a nk-dimensional mesh. Therefore, Proposition 10 gives the Grundy number of an n-dimensional mesh.

Secondly, we give a bound to the parameter for the Cartesian product of odd cycles.

**Proposition 6** Let C3; C5; …; C2k+1 be the cycles of size 3; 5… 2k + 1 respectively. The Grundy number of C3 □C5 □… □C2k+1 is given by:

Γ (C3 □C5 □… □C2k+1) = 2k + 1

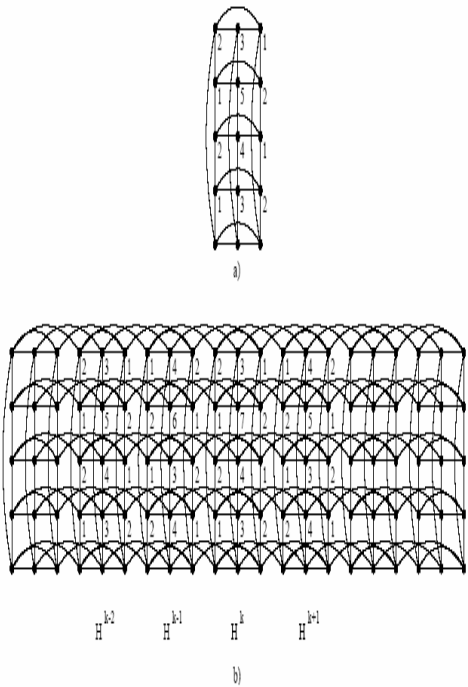

*Figure 4: Partial colorings of a) C3 □C5 and b) C3 □C5 □C7.*

**Remark:** The graph Cn1 □Cn2 □... □Cnk is a nk-dimensional toroidal mesh. So Proposition 10 and Proposition 12 give the Grundy number of a n-dimensional toroidal mesh in two particular cases.

## 4. Generating algorithm

We present a simple recursive algorithm to generate all graphs G with the minimum number of edges such that Γ (G) = k. The main idea is the following: we start from a tree with 2k-1 vertices and we join together some vertices having the same color. By computing all the possible groupings, we find a set of graphs with a Grundy number equals to k. Recursively we start again from each graph of this set while the computation can be done.

## 5. Conclusion

In this article, we first positioned the Grundy number of a graph G compared to other graph parameters (stability number, complement graph of G). Then we presented several bounds and values for the Grundy number of the Cartesian product of two graphs. In particular we studied the Cartesian product of a bipartite graph by a path or a cycle, a bipartite graph by a complete graph and a complete graph by any graph G. Then, we deduced exact values for the Grundy number of a n-dimensional mesh and particular cases of a n-dimensional toroidal mesh.

Thus, for every graph H containing an induced subgraph G presented in this paper, Fact 1 shows that Γ (H) ≥Γ (G) and gives a proper coloring to H.

## ACKNOWLEDGMENT

I think Mr. Brice Effantin for his collaboration